\documentclass[sn-mathphys-num]{sn-jnl}

\usepackage{eqnarray} 
\usepackage{setspace}
\usepackage{dirtytalk}
\usepackage{hyperref}
\hypersetup{
    colorlinks=true,    
    linkcolor=blue,
    filecolor=magenta,      
    urlcolor=cyan,
}
\usepackage{tikz}
\usetikzlibrary{arrows.meta,positioning}
\usepackage{multicol}
\usepackage{multirow}
\usepackage{array}
\usepackage{longtable}
\usepackage{float}
\usepackage{makecell}
\usepackage{natbib}

\usepackage{soul}

\usepackage{subfigure, epsfig}

\usepackage{graphicx}%
\usepackage{multirow}%
\usepackage{amsmath,amssymb,amsfonts}%
\usepackage{amsthm}%
\usepackage{mathrsfs}%
\usepackage[title]{appendix}%
\usepackage{xcolor}%
\usepackage{textcomp}%
\usepackage{manyfoot}%
\usepackage{booktabs}%
\usepackage{algorithm}%
\usepackage{algorithmicx}%
\usepackage{algpseudocode}%
\usepackage{listings}%

\begin{document}

\title[International Vaccine Allocation]{International Vaccine Allocation: An Optimization Framework}


\author[1]{\fnm{Abraham} \sur{Holleran}}\email{mossflower2@gmail.com}

\author*[2]{\fnm{Susan E.} \sur{Martonosi}}\email{martonosi@g.hmc.edu}

\author[3]{\fnm{Michael} \sur{Veatch}}\email{mike.veatch@gordon.edu}

\affil[1]{\orgname{MITRE Corporation}, \orgaddress{\street{202 Burlington Rd.}, \city{Bedford}, \postcode{01730-1420}, \state{MA}, \country{USA}}}

\affil*[2]{\orgdiv{Mathematics Department}, \orgname{Harvey Mudd College}, \orgaddress{\street{301 Platt Blvd.}, \city{Claremont}, \postcode{91711}, \state{CA}, \country{USA}}}

\affil[3]{\orgdiv{Mathematics and Computer Science Department}, \orgname{Gordon College}, \orgaddress{\street{255 Grapevine Rd.}, \city{Wenham}, \postcode{01984}, \state{MA}, \country{USA}}}


\abstract{As observed during the global SARS-CoV-2 (COVID-19) pandemic, high-income countries, such as the United States, may exhibit vaccine nationalism during a pandemic: stockpiling doses of vaccine for their own citizens and being reluctant to distribute doses of the vaccine to lower-income countries.  While many cite moral objections to vaccine nationalism, vaccine inequity during a pandemic could possibly worsen the global effects of the pandemic, including in the high-income countries themselves, through the evolution of new variants of the virus. This paper uses the COVID-19 pandemic as a case study to identify scenarios under which it might be in a high-income nation's own interest to donate vaccine doses to another country before its own population has been fully vaccinated.  We develop an extended SEIR (susceptible-exposed-infectious-recovered) epidemiological model embedded in an optimization framework and examine scenarios involving a single donor and multiple recipient (nondonor) geographic areas.  We find that policies other than donor-first can delay the emergence of a more-contagious variant compared to donor-first, sometimes reducing donor-country deaths in addition to total deaths. Thus, vaccine distribution is not a zero-sum game between donor and nondonor countries: an optimization approach can achieve a dramatic reduction in total deaths with only a small increase  in donor-country deaths.  We also identify realistic scenarios under which the best policy found has a switching form rather than adhering to a strict priority order.  The iterative linear programming approximation approach we develop can help confirm those instances when a priority policy is optimal and, when not optimal, can identify superior policies.  Policies with fewer switching points, which are easier to implement, can be constructed from the best policy found.  This optimization framework can be used to guide equitable vaccine distribution in future pandemics. }




\maketitle

\section*{Highlights}

\begin{itemize}
    \item Vaccine inequity during a pandemic could possibly worsen the global effects of the pandemic, including in nations with ample supply of vaccine, through the evolution of new variants of the virus.
    \item We use an extended SEIR (susceptible-exposed-infectious-recovered) epidemiological model embedded in an optimization framework to examine vaccine distribution scenarios involving a single donor and multiple recipient (nondonor) geographic areas.  
    \item We find that policies other than donor-first can delay the emergence of a more-contagious variant compared to donor-first, sometimes reducing donor-country deaths in addition to total deaths. Thus, vaccine distribution is not a zero-sum game between donor and nondonor countries.
    \item This optimization framework can be used to guide equitable vaccine distribution in future pandemics.
\end{itemize}

\section{Introduction} 



COVID-19 illustrated the huge effect that both pharmaceutical and non-pharmaceutical interventions could have on the course of a pandemic. In this context, a possible contributing factor for the persistence of the pandemic after vaccines first came to market is so-called \textit{vaccine nationalism}.  Initial development of vaccines against COVID-19 was largely funded by high-income nations, including the United States. Meanwhile, access to COVID-19 vaccines in mid- and low-income nations depended largely on vaccine donations from high-income nations \citep{Gavi}. This dependence was a major reason for imbalanced vaccination. By April 2022, high income countries enjoyed a 73\% vaccination rate, and low-income countries achieved only an 11\% vaccination rate \citep{G20Ind}.  


Moral objections to vaccine nationalism during a pandemic abound \citep{HunterEtAl_2022_NEJM, AsundiOLearyBhadelia_2021_CHM, Ghebreyesus_2021_PLOS}. There are also several ways it may harm the nations that practice it. The COVID-19 pandemic illustrated the inability to contain a virus geographically: closing borders proved generally ineffective, only temporarily delaying surges and the spread of variants. Leaving other countries unvaccinated prolongs their surges, resulting in more international transmission and longer travel and trade disruptions. Without vaccination there is also a greater risk of mutation, leading to more contagious variants. For these reasons, it has been argued that vaccine nationalism could undermine a wealthy nation's own interests \citep{AsundiOLearyBhadelia_2021_CHM}. Instead, Emanuel \textit{et al.}'s Fair Priority Model recommends prioritizing countries with the highest transmission rates \citep{Emanuel_Science_2020}.


This paper looks ahead to the next global pandemic and examines the circumstances under which a nation with large vaccine supply might prioritize donating some of their supply to other nations, even before fully vaccinating their own populations, in order to reduce local deaths.  We use data from the COVID-19 pandemic as an exemplar of this generalizeable approach.  We extend an epidemiological disease transmission model to include vaccination and emergence of variants, then embed it within an optimization framework to determine the optimal dynamic allocation of vaccine to geographic regions. The optimization uses iterative linear programming approximation; we also search over static priority policies. We compare four policies: donor-first, nondonor-first, optimized, and optimized with a fairness constraint. To study the emergence of variants, we apply the model in a global context; it could also focus on a group of countries with significant travel between them. The model is high-level and intended for vaccination policy insights, not for short-term forecasting which requires finer data on a smaller geographic area, divided into age/risk groups.

The main contributions of this study are
\begin{itemize}
    \item A compartmental model is developed that includes partial vaccine effectiveness, behavioral or governmental response to surges in cases, and a new model of the emergence of variants.
    \item We show that policies other than donor-first can significantly delay the emergence of a variant compared to donor-first, sometimes reducing donor-country deaths in addition to total deaths.
    \item We demonstrate that vaccine distribution is not a zero-sum game between donor and nondonor countries: policies other than donor-first can achieve dramatic reduction in total deaths with only a small increase (and occasionally even a decrease) in donor-country deaths.
    \item We identify realistic scenarios under which the best policy found has a switching form rather than adhering to a strict priority order. Policies with fewer switching points, which are easier to implement, can be constructed from the best policy found. Policies with this structure could be used to guide vaccine distribution in future pandemics. 
\end{itemize}

The next section presents recent literature on COVID-19 transmission and vaccine allocation policies.  Section \ref{sec:SEIR} develops the SEIR (Susceptible-Exposed-Infected-Recovered) model with vaccination and variants 
and analyzes this model to develop herd immunity thresholds.  In Section \ref{sec:opt} we embed the epidemiological framework into an optimization model to identify optimal vaccine allocations.  
Section \ref{sec:results} describes the scenarios used, presents the results of our approach, and analyzes the structure of the policies yielded by the model.  We conclude and present ideas for future work in Section \ref{sec:conc}.

\section{Literature Review} 
\label{sec:litreview}

The COVID-19 pandemic triggered a wave of research related to prediction of disease trajectory, estimation of disease characteristics, and operational decision-making regarding interventions to control the disease.  There are several surveys of pandemic-related research in the management science and operations management literature  \citep{GuptaEtAl_MSOM_2022, Choi_AOR_2021, JordanEtAl_IEEE_2021, Kaplan_MSOM_2020}.  In particular, Jordan \textit{et al.} note a gap in the literature in the area of decision support \citep{JordanEtAl_IEEE_2021}, which is one purpose of our paper. Although the recent COVID-19 pandemic is a focal point of our literature review, our modeling approach generalizes to future pandemics. 




We focus our review on compartmental models that predict how many people in a population are susceptible (S), exposed (E), infected (I) or recovered (R) from the disease.  Other approaches include agent-based models \citep{SalemMoreno_JContAuto_2022, MahmoodEtAl_JSim_2020, AngelopoulouMykoniatis_JSim_2022, ThompsonWattam_PLOS_2021}, statistical methods \citep{KaplanEtAl_HCMS_2021, LoftiEtAl_AOR_2022, Wang_POMS_2022}, and various other techniques \citep{DuarteEtAl_RiskAnal_2022, KhalilpourazariDoulabi_AOR_2021, MedrekPastuszak_ExpSys_2021}.

\subsection{SEIR Models for Simulation}

In the COVID-19 pandemic, SIR and SEIR models have been used for estimating transmission rate and basic reproduction number of the virus \citep{WangEtAl_PLOS_2020}, validating assumptions about disease transmission and vaccine efficacy \citep{AlgarniEtAl_PeerJCS_2022}, and predicting pandemic dynamics in light of policy, societal behaviors, travel, and social network characteristics \citep{BaggerEtAl_COR_2022, KumarEtAl_AOR_2021_SocMed, ParroEtAl_PLOS_2021, PerakisEtAl_POMS_2022, QianUkkusuri_TransResB_2021}. SEIR models can also be used to simulate the effectiveness of interventions, such as hospital admission policies \citep{ChenKong_POMS_2022}, contact tracing and testing \citep{Sainz-PardoValero_ExpSys_2021}, nonpharmacological interventions such as social distancing, masking, isolating, or quarantining \citep{KumarEtAl_AOR_2021, YuHua_ServiceSci_2021}, and vaccine rollout \citep{KempEtAl_JTheoBio_2021, MakDaiTang_POM_2022}.

These models are useful for short- or medium-term prediction. For longer predictions, SIRS models are used that allow for reinfection by the same or other strains of the virus after immunity wears off \citep{Lazebnik_2022}. Several models include vaccination, including the DELPHI-V model \citep{BertsimasEtAl_HCMS_2021}, \citep{Lazebnik_2022}, and \citep{Shami_2022}. 

Another key feature of high-fidelity models for COVID-19 is partitioning into subpopulations, either by geographic area \citep{Lazebnik_2022}, age and risk groups \citep{VolpertVitalySharma_EcoComp_2021, DolbeaultTurinici_MathMod_2020, GillisEtAl_ORP_2021}, or both \citep{BertsimasEtAl_HCMS_2021}. Volpert, Vitaly, and Sharma examine the effectiveness of vaccination within a heterogeneous population of high transmission and low transmission subpopulations that arise due to characteristics such as age, religious practices, professional experiences, and cultural norms. They find that the effectiveness of vaccination depends on vaccine uptake in each group: achieving a high vaccination rate within the high transmission subpopulation leads to lower overall population rates of infection, but a high vaccination rate only within the low transmission subpopulation serves only to protect the low transmission population against infection. Dolbeault and Turinici also consider the impacts of high- and low-transmission subpopulations in the context of lockdown policies in France. Our work also incorporates interaction between groups, but the groups are countries or regions, and the interaction occurs through the emergence of a variant.

\subsection{Mutation and Multistrain Models}

A key aspect of our model is the emergence of an additional strain, or variant. The simplest multistrain models assume both strains are initially present and interact \citep{Bellamo_2022, KhyarAllali_2020}. Similarly, a second strain can emerge at a fixed time, exogenous to the model \citep{Lazebnik_2022, ShahmanzariEtAl_POMS_2022}. More relevant to our work, Schwarzendahl \textit{et al.} incorporate multiple variants and argue that the average transmission rate is expected to grow linearly with time or with the number of cases \citep{m2}. Our paper assumes that a new variant will emerge after a randomly distributed number of infectious person-days. 


\subsection{Vaccine Allocation Using SEIR Models}

The rapid development of vaccines against COVID-19 raised numerous questions about how  nascent vaccines should be allocated during a pandemic, both among countries and among subpopulations within a given country. Bicher \textit{et al.} consider allocation to risk groups in Austria  \citep{BicherEtAl_PLOS_2022}. Westerink-Duijzer, Schlicher, and Musegaas use cooperative game theory to identify market prices that lead to stable allocations of  influenza vaccines between agencies representing non-interacting regions of the US \citep{Westerink-DjuijzerSchlicherMusegaas_2020}. Gutjahr incorporates equity in health outcomes via a Gini mean absolute difference term to allocate vaccine between interacting geographic regions \citep{Gutjahr_2023}. The model assumes that the entire stockpile is available at the outset of the time horizon, and therefore the model is not time-based; the disease burden as a function of allocation is known \textit{a priori}. Duijzer \textit{et al.} considers allocating influenza vaccine stockpiles to interacting subpopulations, and find that as interaction increases between the subpopulations, disparities caused by inequitable vaccine allocation persist but diminish \citep{DuijzerEtAl_2018_POM}.

Bertsimas, Li, and co-authors develop the DELPHI-V-OPT model  to optimize COVID-19 vaccine allocation to non-interacting geographic areas and age/risk groups in the US (\cite{LiEtAl_OR_2022} and \cite{BertsimasEtAl_HCMS_2021}).  Our optimization method is based on their iterative linear programming approximation.

None of the previous studies consider allocation between countries. Rotesi \textit{et al.} consider interacting countries and examine when it is in a donor country’s best interest to donate vaccines \citep{RotesiEtAl2021}. They demonstrate that it is beneficial to donate vaccines when the donor and recipient countries are close to the herd immunity limit. While we do not model travel directly, we do model the emergence of more contagious variants that appear in the donor country after a time lag. To our knowledge, our paper is the first work that examines the question of optimal vaccine-sharing between countries in the context of geographic interaction and mutation.

\subsection{Other Policy Optimization}

SEIR and other epidemiological models have been used to optimize pandemic mitigation policies other than vaccination. Shahmanzari \textit{et al.} develop a stochastic multiobjective dynamic program that weighs lives lost against economic impacts to compare dynamic and static mitigation strategies to contain COVID-19 disease spread in the context of mutation and vaccination \citep{ShahmanzariEtAl_POMS_2022}. Mitcham and Keisler use a multi-attribute utility decision-making framework that identifies pandemic mitigation strategies which robustly trade-off lives saved, personal liberties, and economic considerations \citep{MitchamKeisler_AOR_2022}. Gillis \textit{et al.} combine a genetic algorithm with an age-stratified SEIR model to identify effective public health responses under various budgetary assumptions \citep{GillisEtAl_ORP_2021}. Salgotra \textit{et al.} intertwine an SEIR model with multi-objective optimization models to examine the tradeoffs between economic costs and health impacts inherent in policies to control COVID-19 transmission \citep{SalgotraEtAl_IEEE_2021}. Vaccine distribution networks have also been optimized  \citep{PanEtAl_AOR_2022, TavanaEtAl_AOR_2021}.

\section{SEIR Model} 
\label{sec:SEIR}

This section presents a Susceptible, Exposed, Infectious, and Recovered (SEIR) model with additional states for vaccinated individuals, making it a \say{SEIR-V} model. There are a small number of geographic areas $a \in \mathcal{A} = \{1,...,n_a\}$ that interact through virus mutation: a more contagious variant emerges after some number of infections and then spreads to the other areas after a fixed time lag. Separate age or risk groups are not considered. 

\subsection{SEIR-V for a Given Variant Emergence}
\label{ss:SEIRwVax}

First we present the epidemic dynamics for a given timing of variant emergence. We make the following assumptions about dynamics.
\begin{itemize}
    \item \textit{Vaccinations:} Vaccinated individuals have lower rates of becoming infected, lower mortality, and are less contagious. Note that, unlike other models, all vaccinated individuals are susceptible to infection.
    \item \textit{Vaccine willingness:} A proportion of the population is willing to be vaccinated. We assume that \say{willingness} is independent of risk of infection (e.g., age or behavior), so that individuals who are willing to be vaccinated are infected at the same rate as unwilling individuals. When there are no more willing susceptible individuals, vaccinations are stopped.
    \item \textit{Infectious time and testing:} A person is infectious until their contagious period ends, they are hospitalized, they are deceased, or they self-isolate or are quarantined due to symptoms or a positive test result. Thus, the rate out of infectious states depends on testing. 
    \item \textit{Mutation:} 
    The infection rate of the new variant is larger than the previous variant \citep{m2}. The timing of when the variant appears and spreads to other areas is addressed in Section \ref{ss:alpha}.
    \item \textit{Behavior:} The infection rate also changes due to social distancing behavior. We assume that the rate is linearly decreasing in the effective number of infectious individuals, i.e., people are more cautious when there is a surge in cases. This feedback loop limits the size of surges. In contrast, many other models treat behavior as fixed or exogenous.
    \item \textit{Reinfection:} Individuals cannot be infected twice. This assumption is reasonable because the time horizon is assumed to be short enough that recovered individuals do not lose their immunity and re-enter the susceptible class.
    \item \textit{Logistics:} We exclude supply chain considerations and assume that vaccines may be immediately reallocated from one country to the other.  
    \item \textit{Time dependence:} All parameters are assumed constant over time except for the amount of vaccine available and the infection rate, which changes due to mutation.
\end{itemize}

\begin{figure}[!ht]
    \centering
\begin{tikzpicture}[node distance=1cm, auto,
    >=Latex, 
    every node/.append style={align=center},
    int/.style={draw, minimum size=1cm, circle}]
   \node [int] (S)             {$S$};
   \node [int, right=of S] (E) {$E$};
   \node [int, right=of E] (I) {$I$};
   \node [int, right=of I] (D) {$D$};
   \node [int, above=of D] (R) {$R$};
    \node [int, above=of S] (SV) {$S^V$};
    \node [int, above=of E] (EV) {$E^V$};
    \node [int, above=of I] (IV) {$I^V$};

   \path[->] (S) edge node {} (E)
   (E) edge node {} (I)
   (I) edge node {} (D)
   (I) edge node {} (R)
   (S) edge node {} (SV)
   (SV) edge node {} (EV)
   (EV) edge node {} (IV)
   (IV) edge node {} (D)
   (IV) edge node {} (R);
\end{tikzpicture}
    \caption{State diagram for the single-area SEIR-V model}
    \label{fig:SEIR-V}
\end{figure}
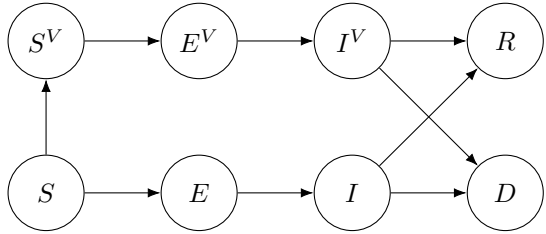

The states of the model are 
diagrammed in Figure \ref{fig:SEIR-V}. These and other variables that vary over time are defined in Table \ref{tab:states}. For example, $S_a(t)$ denotes the number of people in state $S$ in area $a$ after $t$ days. All quantities that depend on area have the subscript $a$; however, it will be suppressed whenever possible. We do not track hospitalizations; instead, the proportion of hospitalized patients that will recover are already counted in $R$ and the rest in $D$.  

{\small \begin{longtable}{| c | l | }
\caption{Variables for each area and time}
\label{tab:states}
    \\ \hline
    Variable & Description \\ \hline
    $S$ & Susceptible unvaccinated (number of people) \\ \hline
    $S^V$ & Susceptible vaccinated \\ \hline
    $E$ & Exposed unvaccinated \\ \hline
    $E^V$ & Exposed vaccinated \\ \hline
    $I$ & Infectious unvaccinated \\ \hline
    $I^V$ & Infectious vaccinated \\ \hline
    $R$ & Recovered \\ \hline
    $D$ & Deceased \\ \hline
    $W$ & In state $S$ and willing to be vaccinated \\ \hline
    $I^E$ & Equivalent number infectious, considering vaccination and behavior. See (\ref{eq:IE}) \\ \hline
    $\beta$ & Infection rate (cases/day/person) \\ \hline    
    $\phi$ & Proportion of new infections that are the variant \\ \hline
 \end{longtable}}
The parameters of the model are listed in Tables \ref{tab:param1} and \ref{tab:param2}. More discussion of their values is given in Section \ref{sec:results} and in Holleran \textit{et al} \citep{HolleranEtAl_Sim}. 

{\small \begin{longtable}{| p{0.1\textwidth} | p{0.15\textwidth} | p{0.65\textwidth} | }
\caption{Parameters of SEIR-V that depend on the area}
\label{tab:param1}
    \\ \hline
    Parameter & Base Value & Description  \\ \hline
    $N_a$ & 100,000 & Initial population  \\ \hline
    $\rho_a$ & 0.78 & Proportion of population willing to be vaccinated  \\ \hline
    $\rho_a^V$ & 0 & Initial proportion of population vaccinated \\ \hline
    $\rho_a^I$   & 0.00072 & Initial new cases per day  as a proportion of the population \\ \hline
    $V_a(t)$ & Policy-dependent & Rate of vaccinations available at time $t$ (people/day)  \\ \hline
    $\gamma^0$ & $1/3.9$ & Rate out of the infectious state without testing (proportion/day)  \\ \hline 
    $\Delta \gamma_a$ & 0.035 \hspace{1in} (donor areas) & Contribution of testing to the rate out of the infectious state (proportion/day)  \\ \hline 
    $\gamma_a$ & $\gamma_a = \gamma^0 + \Delta \gamma_a$ & Rate out of the infectious state (proportion/day)  \\ \hline 
    $\chi_a$ & 1 &Infection multiplier for an area \\ \hline 
 \end{longtable}}

{\small 
\begin{longtable}{| p{0.1\textwidth} | p{0.1\textwidth} | p{0.70\textwidth} | }
\caption{Other parameters of SEIR-V}
\label{tab:param2}
    \\ \hline
    Parameter & Base Value & Description \\ \hline
    $\alpha_0$ & 0.6 & Nominal infection rate, original strain (cases/day/person). \\ \hline 
    $\Delta \alpha$ & 0.6 & Change in infection rate for variant (cases/day/person) 
    \\ \hline
    $I_{\rm{max}}$ & 0.03 & Upper limit on proportion of population infectious, due to behavioral changes \\ \hline 
    $r^I$ & 1/5 & Rate out of the exposed state (proportion/day) \\ \hline
    $p^D$ & 0.014 & Unvaccinated case mortality rate \\ \hline 
    $p_V^D$ & 0.0079 & Vaccinated case mortality rate  \\ \hline 
    $p^e$ & 0.6 & Transmission rate from a vaccinated person as a proportion of rate for an unvaccinated person; $1 - p^e$ is the vaccine effectiveness against transmitting the virus \\ \hline
    $p^r$ & 0.6 & Infection rate for a vaccinated person as a proportion of rate for an unvaccinated person\\ \hline
    $\mu$ & 45,000 & Mean person-days in the infectious state before new variant appears. Only nondonor areas and unvaccinated individuals are counted. \\ \hline
    $CV$ & 1/3 & Coefficient of variation of person-days in the infectious state before new variant appears. \\ \hline
    $L$ & 15 & Lag for the variant to reach other areas (days) \\ \hline
    $T_D$ & 25 & Time for a variant to dominate, i.e., represent half the new cases in an area (days) \\ \hline
    $T$ & 180 & Scenario length (days) \\ \hline
 \end{longtable}
 }

We model the state evolution depicted in Figure \ref{fig:SEIR-V} using discrete time dynamics with an interval of one day ($t = 0,1,\ldots,T$). Given infection rate $\beta(t)$ (which depends on the timing of variant emergence; see Sec. \ref{sec: emergence}), the dynamics can be solved separately for each area.  Vaccination and social distancing behavior change the usual SEIR dynamics in two ways. 

First, the force of infection, $\beta(t)I(t)$, typically used in a SEIR model becomes $\beta(t)I^E(t)$, where $I^E(t)$  represents the \emph{equivalent} number infectious. If we assume no social distancing behavior dynamics, $I^E(t)$ simply accounts for the reduced infectiousness of those in the vaccinated class relative to those in the unvaccinated class:
\begin{equation}
    I^E(t) = I(t) + p^eI^V(t). \label{eq:IE1}
\end{equation} 
To include behavior dynamics, we mimic the approach of \cite{VanOorschotEtAl_2022} and multiply the above by a behavior factor that decreases linearly with this equivalent number infectious: 
\begin{equation}
    I^E(t) = \left( 1 - \frac{I(t) + p^eI^V(t)}{NI_{\rm{max}}} \right)
    [I(t) + p^eI^V(t)]. \label{eq:IE}
\end{equation}
The parameter $I_{\rm{max}} \le 1$ is the proportion $[I(t) + p^eI^V(t)]/N$ of equivalent infections at which $I^E(t)$ drops to zero; it is an upper limit on the proportion of equivalent infections. 

Second, vaccination moves people from $S$ to $S^V$ with nominal rate $V(t)$ given by the vaccine allocation policy\footnote{Throughout this paper, we refer to any schedule of allocations of vaccine to the regions as a \textit{policy}. When a vaccine allocation follows a static ordering of regions, such as donor-first, then nondonor1, nondonor2, etc., we call that a \textit{priority policy}.  When a vaccine allocation distributes vaccine to a particular region, switches to distributing vaccine to another region, and then switches back to distributing vaccine to the previous region, we say that the policy exhibits a \textit{switching form}.  The vaccine allocation that arises as the solution to an optimization problem is referred to as an \textit{optimal policy}.}. However, vaccinations will stop when the number of people in state $S$ at time $t$ who are willing to be vaccinated, $W(t)$, reaches $0$.  Let $t^{*}$ be the last day that $W(t)>0$.   Thus, the rate of vaccinations \emph{administered} is $V^{*}(t)$, which is $V(t)$ prior to time $t^{*}$ and is $0$ after time $t^{*}$.  At time $t=t^{*}$, $V^{*}(t)$ is chosen so that $W(t^{*}+1)=0$.  This is achieved by subtracting from $W(t)$ the number of people moving from state $S$ to state $E$.  We have:
\begin{equation} \label{eq:vstar}
  V^*(t) = \begin{cases}
    V(t), \quad t < t^* \\
    W(t) - \beta(t) \displaystyle \frac{W(t)}{N} I^E(t), \quad t = t^* \\
    0, \quad t > t^*.
   \end{cases} \\
\end{equation}

Using (\ref{eq:IE}) and (\ref{eq:vstar}), the difference equations are
\begin{equation} \label{eq:dyn}
   \begin{array}{rll} 
    S(t+1) & = S(t) - V^*(t) - \beta(t) \displaystyle \frac{S(t)}{N} I^E(t)  
        & t = 0,\ldots, T-1 \\
    S^V(t+1) & = S^V(t) + V^*(t) - p^r\beta(t) \displaystyle \frac{S^V(t)}{N}I^E(t) 
        & t = 0,\ldots, T-1 \\
    E(t+1) & = E(t) + \beta(t) \displaystyle \frac{S(t)}{N} I^E(t) -r^IE(t) 
        & t = 0,\ldots, T-1 \\
    E^V(t+1) & = E^V(t) + p^r\beta(t) \displaystyle \frac{S^V(t)}{N} I^E(t) -r^IE^V(t) 
        & t = 0,\ldots, T-1 \\
    I(t+1) & = I(t) + r^IE(t) - \gamma I(t)
        & t = 0,\ldots, T-1         \\
    I^V(t+1) & = I^V(t) + r^IE^V(t) - \gamma I^V(t)
        & t = 0,\ldots, T-1 \\   
    D(t+1) & = D(t) + \gamma p^DI(t) + \gamma p_V^DI^V(t) 
        & t = 0,\ldots, T-1 \\
    R(t+1) & = R(t) + \gamma(1-p^D)I(t) + \gamma(1-p^D_V)I^V(t) 
        & t = 0,\ldots, T-1 \\
    W(t+1) & = W(t) - \beta(t) \displaystyle \frac{W(t)}{N} I^E(t) - V^*(t) 
        & t = 0,\ldots, T-1. 
    \end{array}
\end{equation}
In these equations, $\beta(t)$ is the infection rate for unvaccinated individuals ($S$), while $p^r\beta(t)$ is the smaller infection rate for vaccinated ($S^V$).  From $E$ (or $E^V$), the rate into $I$ (or $I^V$) is $r^I$ and the rate out of $I$ (or $I^V$) is $\gamma$. The units of these rates are per day, so that in steady state the time spent in the infectious state is $1/\gamma$ days. The total rate out of state $I$ (or $I^V$) is split with proportion $p^D$ (or $p_V^D$) dying and the rest recovering.

\subsection{Emergence of the Variant} \label{ss:alpha}
We propose a new model of when the variant emerges and the resulting infection rate $\beta(t)$. The rate also depends on the area (age distribution, population density, and baseline behavior such as masking and distancing). Behavior dynamics are captured in $I^E(t)$, not in $\beta(t)$; see (\ref{eq:IE}). Initially, a constant mix of strains is assumed to be in all areas, with nominal infection rate $\alpha_0$. The infection rate of the new variant is larger by $\Delta \alpha$. We assume appearance of a variant depends on the number of unvaccinated infectious person-days accumulated over nondonor areas. The assumption that donor areas do not contribute to mutation risk is based on the idea that nondonor areas are largely low-income countries with less immunization against other diseases and more vulnerability to long infections with high viral load. While ideally one could model donor areas as having \emph{lower} mutation risk, rather than none, we are not aware of any source to estimate the relative risk. Sensitivity to this assumption is checked in Section \ref{sec: emergence}. On day $t$, this cumulative number of infectious days is
\begin{equation} \label{eq:Icum} 
I_{\textbf{cum}}(t) = \sum_{a \notin \cal{D}} \sum_{s=0}^{t} I_a(s).
\end{equation}
Making $\beta_a(t)$ depend on (\ref{eq:Icum}) links the areas; their dynamics (\ref{eq:dyn}) can no longer be solved separately.

We model the number of infectious days until the variant emerges as a random variable $X$ with distribution function $F(x)$. If we assume that a variant emerges after a fixed number, $\kappa$, of successive, independent mutations, and that the time elapsed between each mutation is exponentially distributed with mean $c$, then $X$ will follow a gamma distribution with mean $\mu = \kappa c$, and coefficient of variation $CV = 1/\sqrt{\kappa}$.


The model with random $X$ is not tractable for optimization. We approximate the random dynamics with deterministic dynamics by computing the expected infection rates. At each $t$, we will compute the expected $\beta_a(t)$, given $I_{\textbf{cum}}(s), \; s \le t$. Using these $\beta_a(t)$, the dynamics can be solved for $t+1$, obtaining $I_{\textbf{cum}}(t+1)$. Define $Y$ as the day when the variant emerges, using the expected $\beta_a(t)$ to compute $I_{\textbf{cum}}(t)$, which maps $X$ into $Y$. Then
\begin{equation} \label{eq:y}
P(Y = t) = P(I_{\textbf{cum}}(t-1) < X \le I_{\textbf{cum}}(t)) 
  = F(I_{\textbf{cum}}(t)) - F(I_{\textbf{cum}}(t-1)), \quad t = 1, \ldots, T. 
\end{equation}


We also will specify the area in which the variant emerges, $m$, using the deterministic dynamics (using the expected $\beta_a(t)$ to compute $I_{\textbf{cum}}(t)$). In a random model, $m$ is the nondonor area with the largest number of unvaccinated infection-days when the variant emerges; in our deterministic dynamics, we define $m$ as the nondonor area with the largest number of unvaccinated infection-days when $I_{\textbf{cum}}(t)$ reaches $\mu$ (the mean number at which the variant emerges). This occurs on day
\begin{equation} \label{eq:tstar} 
t^* = \min \{t \in \mathbb N: I_{\textbf{cum}}(t) \ge \mu \}. 
\end{equation}


Next we model the expected proportion $\phi(t)$ of new infections in area $m$ that are the variant at $t$. Conditioning on when the variant emerges, we use the sigmoid
\begin{equation} \label{eq:phi} 
\phi(t|Y = s) = \frac{1}{1 + 99^{-[t - (s +T_D)]/T_D}}, \quad t \ge s.
\end{equation}
At $t = s$, $\phi(t) = 0.01$, i.e., when the variant emerges it represents 1\% of new cases. At $t = s + T_D$, $\phi(t) = 0.50$, i.e., $T_D$ is the time until dominance (50\% of new cases). Next we take the expectation over $Y$: 
\begin{equation} \label{eq:phi2} 
\phi(t) = \sum_{s = 1}^{t}\phi(t|Y = s)P(Y = s). 
\end{equation}
Note that for $s > t$, $\phi(t|Y = s) = 0$ so these terms are not needed in the sum.

A deterministic time until the variant emerges is the special case $CV = 0$. While a deterministic model is useful for simulating a given vaccination policy, and we have used it in Holleran \textit{et al} \citep{HolleranEtAl_Sim}, it is not realistic to assume $X = \mu$ is known in advance in an optimization model. Also, in a deterministic model of a single variant, infections after the variant emerges pose no risk, while in a random model infections continue to increase the risk of the variant emerging. For these reasons, we use the random model for optimization. In the deterministic case, the variant emerges when $I_{\textbf{cum}}(t) \ge \mu$ at $t^*$, $Y = t^*$ with probability one, $\phi(t) = 0$ for $t < t^*$ and 1 for $t \ge t^*$, and $\alpha(t) = \alpha_0$ for $t < t^*$ and $\alpha_0 + \Delta \alpha$ for $t \ge t^*$.


The proportion of the variant determines the average nominal infection rate in area $m$, 
\begin{equation} \label{eq:alpha}
\alpha(t) = \alpha_0 + \Delta \alpha \; \phi(t).
\end{equation}
Infection rates in other areas have a lag of $L$ days. Including the behavior factor, the time-varying infection rates are 
\begin{equation} \label{eq:beta}
  \begin{aligned}
    \beta_m(t) & = \alpha(t) \chi_m \\
    \beta_a(t) & = \alpha(\max\{t-L,\,0\} \, )\chi_a \text{   for } a \neq m.
  \end{aligned}
\end{equation}
Figure \ref{fig:alpha_D3_2} in Section \ref{sec:results} shows how $\beta_m$ increases over time. 
This shape seems more realistic than the linear function of $I_{\textbf{cum}}$ used in \cite{m2}.

\subsection{Simulating SEIR-V}
\label{ss:simSEIR}


To solve SEIR-V numerically for a given vaccination policy $V_a(t)$, the difference equations for all areas are solved at successive $t$. The state variables are initialized using the initial proportion vaccinated and initial cases per day. To estimate the exposed states, we use the steady state mean time in these states, $1/r^I$. Multiplying by the new cases per day,
\begin{equation*}
   E(0) + E^V(0)  = \frac{1}{r^I} \rho^I N. 
\end{equation*}
Similarly, for the infectious states
\begin{equation*}
   I(0) + I^V(0) = \frac{1}{\gamma} \rho^I N. 
\end{equation*}
To allocate between vaccinated and unvaccinated states, we use the initial proportion vaccinated and assume cases are only $p^r$ as prevalent among vaccinated individuals. To initialize $W$, $\rho N$ is the number willing to be vaccinated and we assume that those initially in states $E$ and $I$ have the same proportion willing to be vaccinated as $S$. 
Then the initial conditions are 
\begin{align}
    E(0) & = \left( \frac{1 - \rho^V}{p^r \rho^V + 1 - \rho^V} \right)
             \frac{1}{r^I} \rho^I N   \nonumber \\
  E^V(0) & = \left( \frac{p^r \rho^V}{p^r \rho^V + 1 - \rho^V} \right)
             \frac{1}{r^I} \rho^I N   \nonumber\\
    I(0) & = \left( \frac{1 - \rho^V}{p^r \rho^V + 1 - \rho^V} \right)
             \frac{1}{\gamma} \rho^I N                      \label{eq:init}\\    
  I^V(0) & = \left( \frac{p^r \rho^V}{p^r \rho^V + 1 - \rho^V} \right)
             \frac{1}{\gamma} \rho^I N  \nonumber\\                          
  S^V(0) & = \rho^V N - E^V(0) - I^V(0)   \nonumber\\
    S(0) & = N - E(0) - E^V(0) - I(0) - I^V(0) - S^V(0)   \nonumber\\
    W(0) & = \rho N - S^V(0) - E^V(0) - I^V(0) - \rho E(0) - \rho I(0). \nonumber
\end{align}
Note that these values, excluding $W$, sum to $N$, so that $D(0) = R(0) = 0$.

The vaccination policy needs to be modified as shown in (\ref{eq:vstar}) if $W(t)$ reaches 0 (in which case $S(t)$ may also be 0). However, if an area cannot use all of its planned vaccinations, we assume that they can be reallocated to other areas. First, we compute the vaccinations before reallocation as in (\ref{eq:vstar}), now calling them $V^-$ and using the more convenient, equivalent equation 
\begin{equation} \label{eq:v}
    V^-(t) = \min \{W(t) - \beta(t) \frac{W(t)}{N} I^E(t), V(t) \}. 
\end{equation}
The first expression is the number in state $S$ willing to vaccinate before the vaccinations occur.

Now we reallocate the unused vaccinations
\begin{equation}
Q = \sum_{a \in \cal{A}} {[V_a(t) - V_a^-(t)]} \label{eq:q}
\end{equation}
to other areas. First, choose a priority order and arrange the areas in order of descending priority $1,..., n_a$. Starting with $a = 1$, the vaccinations after reallocation in area $a$ are computed as
\begin{equation} \label{eq:v1}
    V_a^*(t) = \min \{W_a(t) - \beta_a(t) \frac{W_a(t)}{N_a} I^E_a(t), V_a^-(t) + Q \}
\end{equation}
and the amount reallocated, $V_a^*(t) - V_a^-(t)$, is subtracted from the available reallocation  Q. This is repeated for all $a$ (or until $Q = 0$).

To summarize, the following sequence of calculations is used to simulate SEIR-V:
\begin{enumerate}
    \item Compute the initial states using (\ref{eq:init}) and $\beta_a(0) = \alpha_0 \chi_a$.
    \item For $t = 0, \ldots, T - 1$, solve the difference equations (\ref{eq:dyn}) along with (\ref{eq:IE}), (\ref{eq:Icum})-(\ref{eq:beta}) for $\beta$ and (\ref{eq:v})-(\ref{eq:v1}) for $V^*$ for all areas at the current $t$. \\
\end{enumerate}

\subsection{Herd Immunity}
\label{ss:herd}
This section presents herd immunity conditions for the continuous time version of our model that will be useful in interpreting the numerical results. Herd immunity is defined as stability to a small injection of infections, from an initial state with no infections. We can remove the states $E,E^V$ for the purpose of stability analysis and assume that $I(t)$ and $I^V(t)$ are infinitesimal, so that $\beta$, $S(t)$ and $S^V(t)$ are constant over the time scale of stability analysis. Because they are infinitesimal, we can also use (\ref{eq:IE1}), ignoring the behavior dynamics in (\ref{eq:IE}). Let $\dot{I}$, etc. denote derivatives. From (\ref{eq:dyn}) and (\ref{eq:IE}), 
\begin{align*}
\dot{I} & = \beta \frac{S}{N}(I + p^e I^V) - \gamma I \\
\dot{I}^V & = p^r \beta \frac{S^V}{N}(I + p^e I^V) - \gamma I^V.
\end{align*} 
If no vaccinations occur, $S^V = I^V = 0$ and the stability condition for the unvaccinated group is $\dot{I} < 0$, or
\begin{equation}\label{eq:SN}
\frac{S}{N} < \frac{\gamma}{\beta} = \frac{1}{R_0},
\end{equation} 
where $R_0$ is the unvaccinated basic reproduction number. Since $\beta_a(t)$ varies after the variant emerges, so does $R_0$.  If all susceptible individuals are vaccinated, then $S = I = 0$, and the stability condition is $\dot{I}^V < 0$, or
\begin{equation}\label{eq:SVN}
\frac{S^V}{N} < \frac{\gamma}{p^rp^e\beta} = \frac{1}{R_0^V},
\end{equation} 
where $R_0^V$ is the vaccinated basic reproduction number. This is weaker than (\ref{eq:SN}), since $p^r,p^e < 1$. Also define the \emph{critical proportion} as the proportion that must \emph{not} be susceptible in order to reach herd immunity. With no vaccination, the critical proportion is $1 - S/N = 1 - 1/R_0$, and similarly when all are vaccinated.  

Stability conditions for the general case where $S,S^V > 0$ are more complex; stability of a similar model is analyzed in \citep{Fudolig_2020}. The local stability condition for a system of differential equations is that the eigenvalues of the Jacobian all have negative real components. The Jacobian is
\begin{equation*}
J =  
\begin{bmatrix}
\displaystyle \frac{\partial \dot{I}}{\partial I} & \displaystyle \frac{\partial \dot{I}^V}{\partial I} \\[0.1in]
\displaystyle \frac{\partial \dot{I}}{\partial I^V} & \displaystyle \frac{\partial \dot{I}^V}{\partial I^V}
\end{bmatrix}
= 
\begin{bmatrix}
\beta S/N - \gamma & p^r\beta S^V/N \\[0.1in]
p^e\beta S/N & \: p^r p^e\beta S^V/N - \gamma
\end{bmatrix}
\end{equation*} 
with eigenvalues $-1$ and
\[ \frac{p^r p^e\beta S^V}{\gamma N} + \frac{\beta S}{\gamma N} - 1. \] 
Thus, the stability condition is 
\begin{equation}\label{eq:S_tot}
\frac{S + p^e p^r S^V}{N} < \frac{\gamma}{\beta}.
\end{equation} 

Another quantity of interest is the proportion of cases that are vaccinated. Let $\psi(t) = I^V(t)/I(t)$ be the ratio of vaccinated to unvaccinated infectious at time $t$. For a given $S$ and $S^V$, equilibrium for $\psi$ occurs when
\[ \frac{\dot{I}^V/I^V}{\dot{I}/I} =  \frac{p^r\beta (1/\psi + p^e)S^V/N - \gamma}
                                     {\beta (1 + p^e \psi)S/N - \gamma}         
                                   = 1 \]
with solution
\[ \psi = \frac{p^r S^V}{S}. \]
Further, at equilibrium the relative risk of a new infection (vaccinated to unvaccinated) is
\[ \frac{I^V/S^V}{I/S} =  \psi \frac{S}{S^V} = p^r \]
and the proportion of new cases that are vaccinated is
\[ \frac{I^V}{I + I^V} =  \frac{p^rS^V}{p^rS^V + S}. \]
We conjecture that the equilibrium for $\psi$ is stable, so that the proportion of cumulative cases that are vaccinated will tend to track this value even as $S$ and $S^V$ change slowly over time.  



\section{Optimization Framework}
\label{sec:opt}

In this section we formulate a model that allocates the vaccines available each day to areas in order to minimize deaths. Let $\mathcal{D}$ be the set of donor areas. The optimization problem, called SEIR-OPT, is
{\small
\begin{equationarray}{rllll}
 \multicolumn{3}{l}{\text{min   } \displaystyle \sum_{a\in \mathcal{D}} D_a(T) 
    + \nu \sum_{a\notin \mathcal{D}} D_a(T)} & & \label{eq:deaths} \\
    \text{s.t.  } \displaystyle \sum_{a\in \mathcal{A}}V_a(t) &\leq & B(t)  
        & t = 0,\ldots, T-1 & \label{eq:budget}\\
     W_a(t+1) & = & W_a(t) - \beta_a(t) \frac{W_a(t)}{N_a}I^E_a(t) - V_a(t)
        & t = 0,\ldots, T-1, & a \in A \label{eq:W} \\
    S_a(t+1) & = & S_a(t)  -V_a(t) - \beta_a(t) \frac{S_a(t)}{N_a}I^E_a(t)& t = 0,\ldots, T-1,& a \in A \label{eq:S}\\
    S^V_a(t+1) & = & S^V_a(t) + V_a(t) - p^r\beta_a(t) \frac{S^V_a(t)}{N_a}I^E_a(t)& t = 0,\ldots, T-1,& a \in A \\
    E_a(t+1) & = & E_a(t) + \beta_a(t) \frac{S_a(t)}{N_a}I^E_a(t) -r^IE_a(t) & t = 0,\ldots, T-1,& a \in A\\
    E^V_a(t+1) & = & E^V_a(t) + p^r \beta_a(t) \frac{S^V_a(t)}{N_a}I^E_a(t) -r^IE^V_a(t) & t = 0,\ldots, T-1,& a \in A \label{eq:EV}\\
    I_a(t+1) & = & I_a(t) + r^IE_a(t) - \gamma_aI_a(t)& t = 0,\ldots, T-1,& a \in A\\
    I^V_a(t+1) & = & I^V_a(t) + r^IE^V_a(t) - \gamma_aI^V_a(t)& t = 0,\ldots, T-1,& a \in A\\  
    D_a(t+1) & = & D_a(t) + \gamma_ap^DI_a(t) + \gamma_ap_V^DI_a^V(t) & t = 0,\ldots, T-1,& a \in A\\
    R_a(t+1) & = & R_a(t) + \gamma_a(1-p^D)I_a(t) + \gamma_a(1-p^D_V)I^V_a(t)& t = 0,\ldots, T-1,& a \in A \label{eq:R}\\
    I^E_a(t) & = & \left( 1 - \frac{I_a(t) + p^eI_a^V(t)}{NI_{\rm{max}}} \right) [I_a(t) + p^eI_a^V(t)] & t = 0,\ldots, T-1,& a \in A \label{eq:IEopt}\\
    \multicolumn{3}{l}{W_a(t), S_a(t), E_a(t), I_a(t), S^V_a(t), E_a^V(t), I_a^V(t), V_a(t) \geq 0} & t = 0,\ldots, T, & a \in A. \nonumber
\end{equationarray}}
The variables in SEIR-OPT are all of the state variables, $V$, and $I^E$ at times $t = 1,\ldots, T$. Their initial values at $t=0$ are computed in (\ref{eq:init}). Also, $\beta$ is a variable because the timing of variant emergence depends on $I$; see Section \ref{ss:alpha}. Nondonor deaths are given a weight $0 \le \nu \le 1$ in the objective. We consider deaths in the donor areas (self-interest) by setting $\nu = 0$ or all areas (altruism) by setting $\nu = 1$. Constraint (\ref{eq:budget}) limits vaccinations to the budget of $B(t)$ doses available for day $t$,
(\ref{eq:W})-(\ref{eq:R}) are the difference equations, and (\ref{eq:IEopt}) computes $I^E$. There is no need to adjust $V$ using (\ref{eq:v}) - (\ref{eq:v1}) because the constraints make $V$ feasible. 

One cannot directly solve SEIR-OPT due to the complex dependence of $\beta$ on $I$ through the timing of the variant. 
Instead, we iteratively solve a \say{Lagrangian} problem. As a surrogate for the cost of the variant emerging earlier, 
we use the unvaccinated, nondonor infectious-days to construct a \say{Lagrangian} term. Weighting infections by the time remaining in the scenario gives somewhat better numerical results. The new objective is
\begin{equation} \label{eq:lagr}
 \text{min    } \sum_{a\in \mathcal{D}} D_a(T) + \nu \sum_{a\notin \mathcal{D}} D_a(T) + \lambda \sum_{a \notin \mathcal{D}} \sum_{t=1}^{T}{I_a(t)(T - t)}.
\end{equation}
Instead of letting 
$\beta$ depend on $I$ in (\ref{eq:beta}), we fix their values. For the first iteration we use the value from an initial policy, then update it each iteration using the policy found in the previous iteration. 

Even with constant $\beta$, the constraints contain cubic terms, such as $W_a(t)I^E_a(t)$ (recall that $I^E_a(t)$ is quadratic), in (\ref{eq:W}-\ref{eq:EV}). These terms result in nonconvex constraints, making the Lagrangian problem very difficult to solve. If the behavior dynamics in (\ref{eq:IE}) are removed, using (\ref{eq:IE1}) instead, the constraints are nonconvex quadratics. Bertsimas, et. al \citep{a1} solve a similar problem by solving a sequence of linear approximations. At each iteration, given the current vaccine allocation $V$, (\ref{eq:W}) - (\ref{eq:IEopt}) are solved to get the current infectious population estimates $\hat{I}_a(t)$, $\hat{I}^V_a(t)$, and $\hat{I^E}_a(t)$, as well as 
$\hat{\beta}_a(t)$.  Then we replace the variables $I^E$ and $\beta$ in (\ref{eq:W}) - (\ref{eq:EV}) with the constants $\hat{I^E}$ and $\hat{\beta}$. To keep the linearization error from being too large, we add linearized regularization constraints to prevent $I^E$ from differing too much from $\hat{I^E}$:
\begin{equation} \label{eq:i}
    \mid G_a(t)[I_a(t) + p^eI^V_a(t)]- \hat{I^E}_a(t) \mid \leq \epsilon \quad t = 1, \ldots, T-1, \quad a \in A.\\
\end{equation}
Without behavior dynamics, $G_a(t) = 1$ to match (\ref{eq:IE1}). With behavior, 
\begin{equation*}
G_a(t) = 1 - \frac{\hat{I}_a(t) + p^e \hat{I}_a^V(t)}{N_a I_{\rm{max}}}
\end{equation*}
to match (\ref{eq:IE}). 
Here, $\epsilon$ is an exploration tolerance that is updated at each iteration by a multiplicative factor $\phi < 1$. To summarize, the approximation is a linear program (LP) that differs from SEIR-OPT in the objective (\ref{eq:lagr}), fixed $\beta = \hat{\beta}$, fixed $I^E = \hat{I^E}$, and added constraints (\ref{eq:i}). 

The iterative approach to the Lagrangian problem solves this LP to find the best solution $V^{\text{new}}$ that gives effective infections $I^E$ that are close to $\hat{I^E}$. The current solution $V$ is updated and the difference equations solved to find $\hat{I^E}_a(t)$, 
and $\hat{\beta}_a(t)$. We call this step \say{Simulate}. The process is repeated until the objective function converges. Because the Lagrangian problem is not convex, this algorithm is not guaranteed to converge to a global minimum; see Section \ref{sec:convergence}. To start the algorithm, an initial policy $V$ is chosen and simulated. We use static priority policies that allocate vaccine in priority order.

Algorithm \ref{alg} describes how we solve SEIR-OPT. It has an outer loop that updates $\lambda$ and an inner loop that updates $I^E$ and $\beta$. The inner loop seeks to solve the Lagrangian problem. Let $L(\lambda)$ denote the optimal value of the Lagrangian (\ref{eq:lagr}) and $z(\lambda)$ be the corresponding SEIR-OPT objective (\ref{eq:deaths}). Each iteration $j$ of the inner loop solves the LP to find $V$, then simulates $V$ to obtain a valid solution $V^*$ to the difference equations. Let $L^{(j)}$ be the Lagrangian (\ref{eq:lagr}), evaluated for this $V^*$ and $\lambda$, and $z^{(j)}$ be the corresponding SEIR-OPT objective (\ref{eq:deaths}). Successive LPs may not be improving, so the best $L^{(j)}$ for the current $\lambda$ is stored in $L^{\min}$. Similarly, it may not terminate at the LP with the best SEIR-OPT objective (\ref{eq:deaths}), so the best $z^{(j)}$ is stored in $z^{\text{opt}}$ and the corresponding best simulated policy in $V^{\text{opt}}$. The stopping criterion shown for the inner loop is that $L^{(j)}$ changes by less than $\delta$ in one iteration; however, because $L^{(j)}$ does not always improve between iterations, instead of $\delta$ we used an iteration limit and checked convergence by computing the maximum change in the LP variables $V_a(t)$.

The outer loop searches for the minimum of $z(\lambda)$. Since $z$ is the SEIR-OPT objective function, this minimum occurs at the best solution to SEIR-OPT found by the Lagrangian problem.  
We use a global search method, using geometrically spaced $\lambda$'s over an initial range, selecting the best interval(s), and iterating with a finer partition for $\lambda$. The parameters used in the algorithm are listed in Table \ref{tab:param3}. 

\begin{algorithm}[!ht]
\caption{Algorithm for SEIR-OPT} \label{alg}
\begin{algorithmic}
\State Initialization: $i \leftarrow 0, \lambda^{(1)} \leftarrow \lambda.$ Choose a priority policy.
\State Simulate and use it to update $t_n$, $\beta_a(t)$, $\hat{I^E} \leftarrow I^E$, $V^{\text{opt}} \leftarrow V^*$, and $z^{\text{opt}}$ from (\ref{eq:deaths}).
\While{searching over $\lambda$}
\State $i \leftarrow i + 1, j \leftarrow 0, \epsilon \leftarrow \epsilon_0$, $L^{\min} \leftarrow \infty$
\While{$|L^{(j)} - L^{(j-1)}| > \delta$ or $j<2$}
\State $j \leftarrow j + 1$
\State Solve the LP using $t_n$, $m$, $\hat{I^E}$, $\lambda^{(i)}$ and $\epsilon$, giving $L^{(j)}$ and $V$.
\State Simulate $V$ to get $V^{\text{opt}}$ and update $t_n$, $\beta_a(t)$, $\hat{I^E} \leftarrow I^E$, $z^{(j)}$ from (\ref{eq:deaths}), and $L^{(j)}$ from (\ref{eq:lagr}).
\State If $L^{(j)} < L^{\min}$, update $L^{\min} \leftarrow  L^{(j)}$.
\State If $z^{(j)} < z^{\text{opt}}$, update $z^{\text{opt}} \leftarrow  z^{(j)}$, $V^{\text{opt}} \leftarrow  V^*$.
\State $\epsilon \leftarrow \phi \epsilon$
\EndWhile
\State $z(\lambda^{(i)}) \leftarrow z^{(j)}$
\State If continuing search over $\lambda$, set $\lambda^{(i+1)}$. 
\EndWhile
\State \textbf{output:} Vaccine allocation $V^{\text{opt}}$ and donor deaths $z^{\text{opt}}$.
\end{algorithmic}
\end{algorithm}

\begin{table}[!h] 
\caption{Parameters for optimization model}
\label{tab:param3}
\begin{tabular}{| l | l | l |}
    \hline
    Parameter & Base Value  & Description  \\ \hline
    $B(t)$ & 1500 & Vaccine doses available on day $t$ \\ \hline
    $\nu$ & 0 & Objective function weight on nondonor deaths \\ \hline
    $\lambda$ & $10^{-6}$ to $10^{-4}$ & Range of Lagrange multiplier in search \\ \hline
    $\epsilon_0$ & 500 & Initial exploration tolerance for LP \\ \hline        
    $\phi$ & 0.8 & Exploration convergence parameter for LP \\ \hline
\end{tabular}
\end{table}

\section{Results}
\label{sec:results}


All numerical tests used Gurobi 11.0 on a laptop with a 1.70GHz processor, 16 GB of RAM, and four physical cores. We focus on scenarios with more than two areas, where more interesting optimal policies were found. Baseline parameter values are shown in Tables \ref{tab:param1} and \ref{tab:param2}. These parameter estimates are based on COVID-19 data and are derived in Holleran \textit{et al} \citep{HolleranEtAl_Sim}, except where noted here:

\begin{itemize}

\item \textbf{$N_a$ (initial population)}:  We assume the donor population is twice as large as in each nondonor area. For three areas, then, the donor comprises 50\%. This approximates a global scenario with the main producers of COVID-19 vaccine (China, the European Union, India, the Unites States, and Russia) as the donor area, since they represent 47\% of global population, and other countries divided into two areas with equal populations (roughly, they could be Africa and all others). The scenario with 10 areas assumes more geographic isolation. It could apply to a pandemic model where China and India are sufficiently isolated to be excluded, the European Union and the United States are the donor area, and other nations are grouped into nine nondonor areas. The donor population is arbitrarily set to 100,000, i.e., the donor deaths reported are per 100,000.


\item \textbf{$\rho_a^I$ (initial rate of new infections)}:  $\rho_a^I$ is multiplied by the average duration of infectiousness to obtain the initial prevalence. Thus, $\rho_a^I = (\textrm{prevalence}) r_a^d$. One COVID-19 study estimates a prevalence in the U.S. of $1.4\%$ on December 31, 2020 \citep{PLOS_US_prev}. Converting, $\rho_a^I = 0.014 r_a^d = 0.0036$. However, cases were higher in January 2021 and much higher in January 2022, so we use $\rho_a^I = 0.0072$ as a baseline in all areas.



\item \textbf{Person-days of infection before appearance of new variant:} The mean $\mu = 45,000$ is set so that the variant risk increases in the middle of the 180 day scenario; see Figure \ref{fig:alpha_D3_2}. The coefficient of variation is set to $CV = 1/3$, reasoning that the number of independent mutations is on the order of $\kappa \approx 10$ and $CV = 1/\sqrt{\kappa}$, as described in Section \ref{ss:alpha}.


\end{itemize}

\noindent Table \ref{tab:scenarios} lists parameter changes for the scenarios referenced in the following sections. Scenario 3.1 uses the baseline values, so it is based on COVID-19 data when available.

For Scenario 3.1, policies other than donor-first, and policies that do not use static priorities, are beneficial. We then explored other scenarios to find when this benefit was most significant, resulting in those in Table \ref{tab:scenarios}. In the other scenarios, $p_V^D = p^D = 0.014$, meaning that vaccination does not reduce the case mortality rate. While this is not realistic for COVID-19, it may be for some viruses.


\begin{table}[!h] 
\caption{Scenarios for optimization. Values from baseline scenario are in bold.}
\label{tab:scenarios}
\begin{tabular}{| l | c | c | c | c |}
\hline 
 Scenario & 3.1 & 3.2 &  4.1 & 10.1 \\ \hline\hline
  Areas & 3 & 3 &  4 & 10 \\ \hline
 Population $N_a$ (1000's) & \textbf{100, 50, 50} & \textbf{100, 50, 50} & \textbf{100, 50, 50, 50} 
            & \textbf{100, 50,...,50} \\ \hline
 Initial new cases & \textbf{0.00072}, & 0.0018, & 0.0018, & 0.001, \\
 per person & \textbf{0.00072}, & \textbf{0.00072}, & \textbf{0.00072},... & 0.002, 0.0018,... \\
 per day $\rho^I_a$ & \textbf{0.00072} & \textbf{0.00072} & \textbf{0.00072} & 0.0006, 0.0004 \\ \hline
Infection multiplier $\chi_a$  & \textbf{1, 1, 1} & \textbf{1, 1, 1} & \textbf{1, 1, 1, 1} & 1.5, 1,...,1 \\ \hline
Vaccinated mortality rate $p^D_V$ & \textbf{0.0079} & 0.014 & 0.014 & 0.014 \\ \hline
Mean infectious days & & & & \\
before variant $\mu$ (1000's) & 55 & 50 & 75 & 300 \\ \hline
Coefficient of variation & & & & \\
of infectious days & & & & \\
before variant $CV$ & \textbf{1/3} & 0.71 & \textbf{1/3} & \textbf{1/3} \\ \hline
Vaccine doses available & & & & \\
each day $B(t)$      & \textbf{1500} & \textbf{1500} & 2000 & 3000 \\ \hline
\end{tabular}
\end{table}

\subsection{Convergence of the Algorithm} \label{sec:convergence}

In this section we test the convergence of the algorithm. Three questions are of interest:
\begin{enumerate}
    \item Is the solution of the Lagrangian problem (\ref{eq:lagr}) for some $\lambda$ near-optimal for the original problem? That is, is the surrogate for the Lagrangian adequate?
    \item Does the global search find (nearly) the best value of $\lambda$?
    \item Does the iterative LP approximation converge to a near-optimal solution of the Lagrangian problem, independent of the initial policy?
\end{enumerate}
We cannot definitively answer the first question because the original problem is intractable. Thus, we refer to the best policy found by the algorithm, rather than the optimal policy. However, after extensive testing on scenarios with three areas, it is encouraging that when one of the priority policies is much better than the others, our algorithm converges to that policy and that when priority policies are nearly tied, our algorithm usually finds a policy noticeably better than the best priority policy.

\begin{figure}[!ht]
    \centering
    \includegraphics[width=3in]{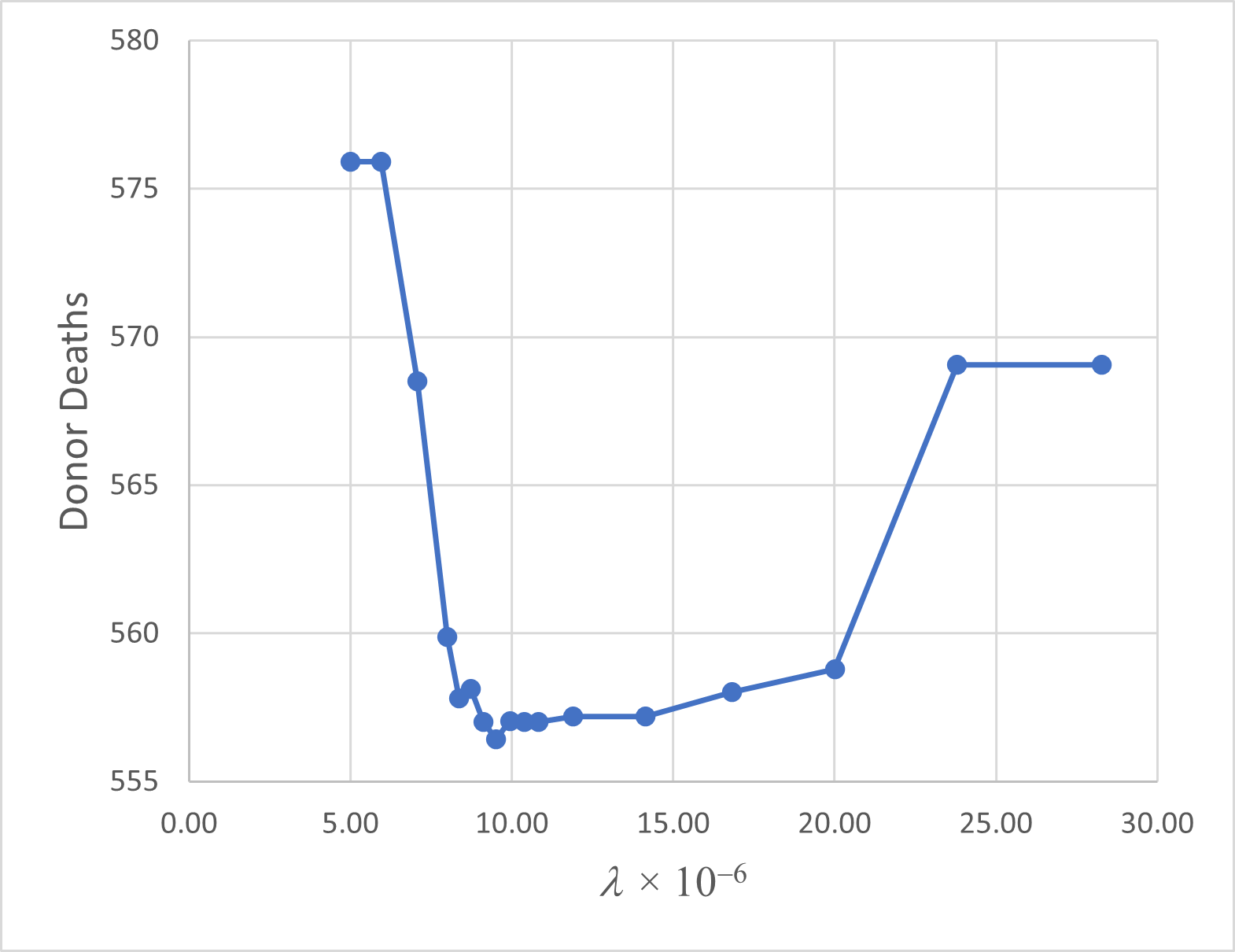}
    \caption{Best Donor Deaths vs. $\lambda$ (Scenario 3.2)}
    \label{fig:lambda}
\end{figure}

Figure \ref{fig:lambda} is a representative example of how the objective of the best policy found varies with $\lambda$. The policy structure generally varies as expected with $\lambda$: for small $\lambda$, a donor-first policy is used and as $\lambda$ increases, donor vaccinations begin later. At the largest two values of $\lambda$ graphed, they start on day 47 and the policy is essentially donor-last (the last 140 vaccinations in the nondonor areas occur on day 85 at the end of all vaccinations, but this appears to be due to not finding the exact optimum).    

This graph is not quite unimin: there are two local minima with slightly larger objective function values between them. It also has a flat spot to the right of the global minimum, where the objective and the policy are essentially constant. These features necessitate a global search. Also, the fact that the non-monotone behavior occurs close to the global minimum motivates our use of a search algorithm that iteratively refines the grid near the best $\lambda$ found. The slightly non-monotone behavior appears to be an artifact. The iterative LP convergence is approximate, so the best objective value found is also approximate. The solution for a given $\lambda$ can also depend slightly on the previous solution, as its policy is used for initialization. These tests suggest that the search over $\lambda$ is finding the global minimum, or very nearly so.

We tested the convergence of the iterative LP approximation for $\epsilon_0 = 50$ to 1000 and $\phi = 0.6$ to 0.9 for Scenario 3.2. We found that when these step size parameters were large, the algorithm gets close to the best policy in very few iterations. However, after that the objective and policy oscillate with a period of two, presumably because it overshoots the reallocation to other areas (for more than three areas the period might be different). For smaller tolerances, the algorithm takes smaller steps, usually requires 10-30 iterations to reach its best policy, and again oscillates once it gets close to the best policy. We also measured the maximum change in $I^E(t)$, which is restricted by the regularization constraint (\ref{eq:i}). We found that when the best policy is complex, this change remains large, i.e., the disease trajectory continues to change significantly between iterations, until it is limited by (\ref{eq:i}). When the best policy is a priority policy, the policy converges quickly and so does the change in $I^E(t)$. We chose $\epsilon_0 = 500$ and $\phi = 0.8$ because they found a slightly better policy than the other values tested; they also performed well in slight variants of Scenario 3.2. This is surprising, given the oscillating behavior. For simplicity, we used these values in all scenarios, rather than varying them for each scenario.

To check sensitivity to the initial policy, we tried every priority policy for Scenario 3.2. All objective values found were within 0.04\%. The number of iterations required to find the best policy was also similar (14 or 15) for every initial policy, presumably because the optimal policy is far from a static priority. 

These tests suggest that, while the algorithm does not find the exact optimal policy, it consistently finds a better, more complex policy. The parameter values, initial policy, and iteration limits used appear adequate to obtain the best, or very close to best, policy the algorithm can find. The structure of the policies found, discussed in the next section, also seem reasonable, suggesting that they are close to optimal.

\subsection{Policy Structure}
\label{sec: emergence}

\begin{figure}[!ht]
    \centering
    \includegraphics[width=\textwidth]{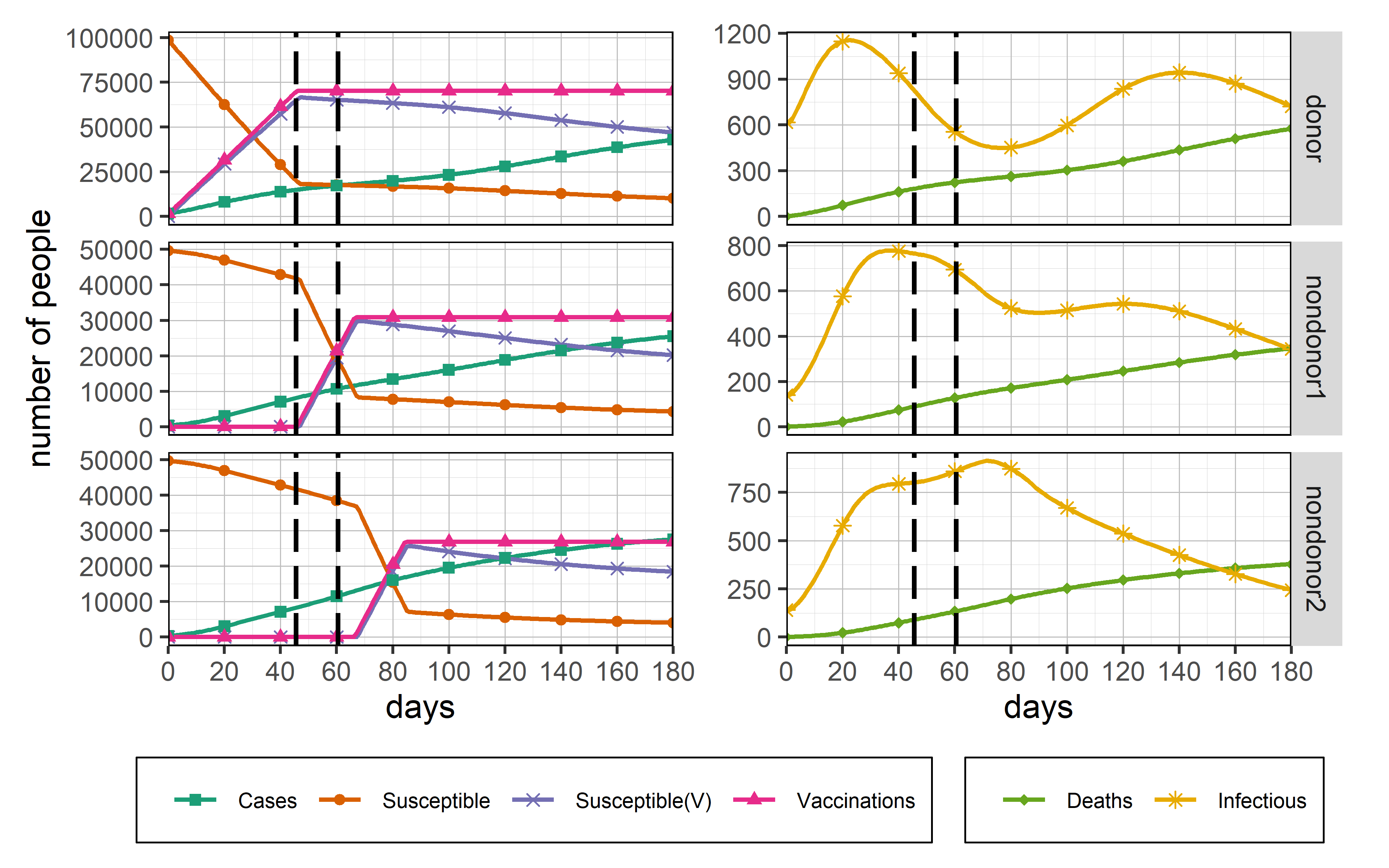}
    \caption{Scenario 3.2, donor-first policy}
    \label{fig:time_D3_2}
\end{figure}

\begin{figure}[!ht]
    \centering
    \includegraphics[width=\textwidth]{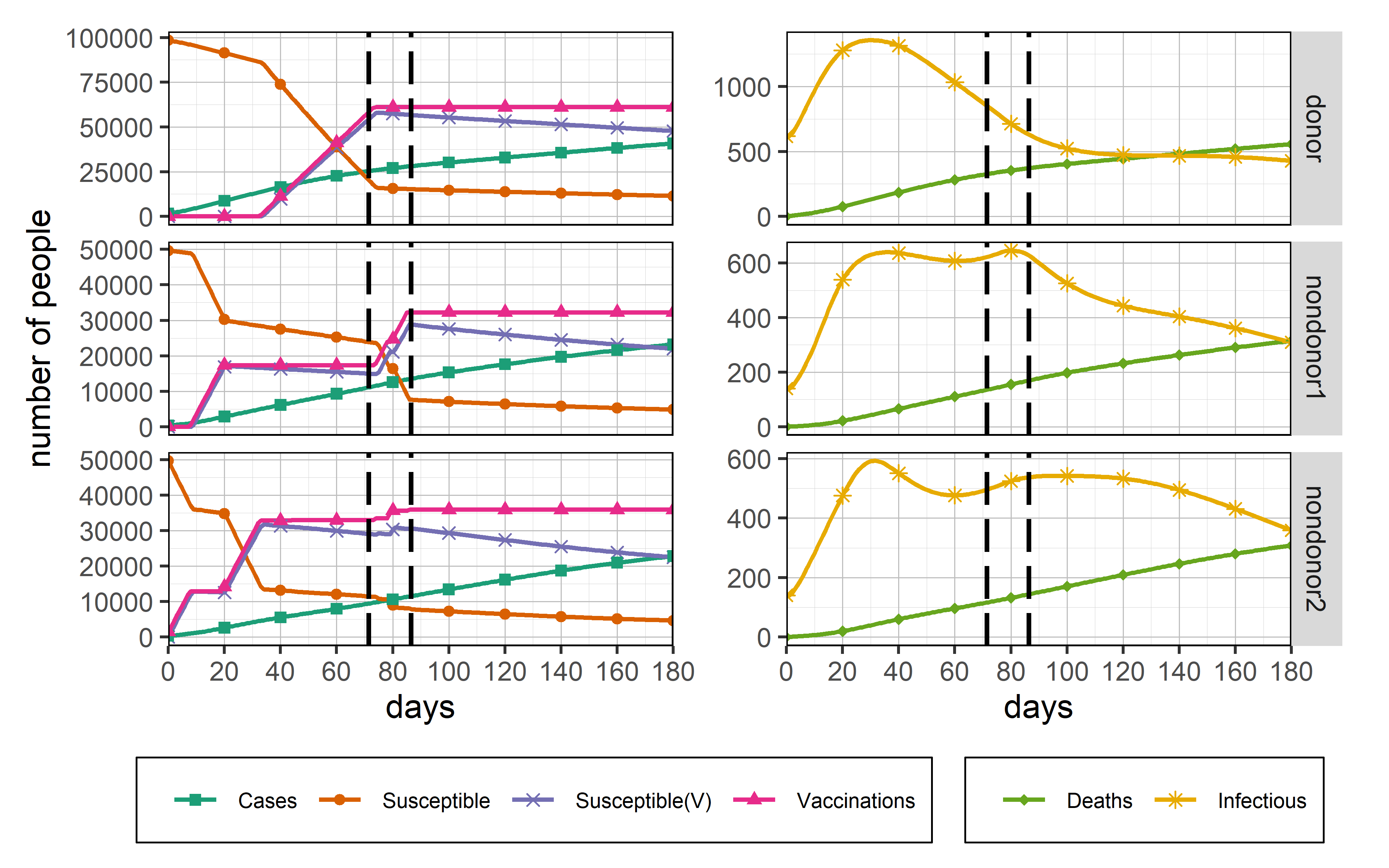}
    \caption{Scenario 3.2, best policy found}
    \label{fig:time_opt3_2}
\end{figure}

\begin{figure}[!ht]
    \centering
    \includegraphics[width=4in]{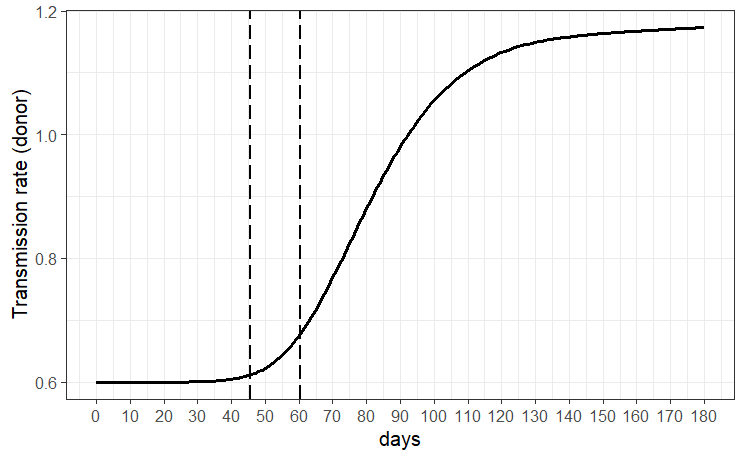}
    \caption{Transmission rate for donor area, Scenario 3.2, donor-first policy}
    \label{fig:alpha_D3_2}
\end{figure}

Results for Scenario 3.2 under the donor-first and best policy found are shown in Figures \ref{fig:time_D3_2} and \ref{fig:time_opt3_2}. Because the nondonor areas are identical, we arbitrarily give nondonor1 priority over nondonor2. Cumulative vaccinations, cases, and deaths, as well as the current susceptible and infectious ($I(t)+I^v(t)$) populations are shown by area. For both policies, the variant emerges in area nondonor2. The two vertical lines indicate the expected time when the variant arrives in this area and then the other areas (more precisely, the time when the number of cases reaches the mean number of cases at which the variant emerges). Figure \ref{fig:alpha_D3_2} shows how, for the donor-first policy, the transmission rate increases from that of the current variant (0.6) to almost that of the new variant (1.2) as the risk of the variant appearing increases. The transmission rate increases more quickly than for the best policy because there are more cases in the nondonor unvaccinated populations.

We also observe in Figure \ref{fig:time_D3_2} the emergence of a second wave of infections, which can be understood from the herd immunity critical proportions in Table \ref{tab:herd}.  For the donor-first policy, the donor area reaches its vaccination limit of 78\% at day 46. However, by day 82 the variant is 50\% of cases in the donor area and infections rebound, forming a second wave.  Before the variant, herd immunity in a fully unvaccinated donor area would require a critical (protected) proportion of 51\%, whereas a fully vaccinated donor area would require a critical proportion of 0\%.  Thus, the actual critical proportion prior to variant emergence is within this range. This is reached before roughly day 20, when cases are 10\% and susceptible vaccinated are 30\%. However, as the transmission rate increases the population moves out of herd immunity. For example, in the donor area on day 82 the transmission rate is $\beta = 0.9$, there are 19,482 cumulative cases, 16,636 susceptible, and 63,198 susceptible vaccinated. The left side of  (\ref{eq:S_tot}) equals 0.394 and the right side is 0.409, barely meeting herd immunity; just after this the donor area loses herd immunity as $\beta$ increases. The donor area moves back into herd immunity near the end of the scenario as cases build.    


For the best policy, the donor area has a larger first wave, but even without vaccination, it is moderated by behavior (fewer contacts). The variant arrives later, having little impact on transmission rates until all areas are vaccinated, and there is only a very small second wave. Comparing the nondonor areas in Figures \ref{fig:time_D3_2} and \ref{fig:time_opt3_2}, the first surge is much smaller under the best policy, resulting in the variant arriving much later. Under the best policy, total cases are only slightly reduced but are delayed significantly, both in donor and nondonor areas.

The best policy has a switching form. As can be seen from the vaccinations plotted in Figure \ref{fig:time_opt3_2}, it switches between the nondonor areas, then switches to the donor area for days 33-73 until its vaccination limit is reached, then switches back to nondonor1. Some additional switching occurs, however, it amounts to less than one day's vaccination budget and may be due to the algorithm not finding the exact optimal. As shown in Section \ref{sec:optimization}, the best policy found is significantly better than any of the priority policies.

\begin{table}[!h] 
\caption{Herd immunity critical proportions (see Section \ref{ss:herd})}
\label{tab:herd}
\begin{tabular}{| l | c c | c c | cc|}
\hline 
& \multicolumn{6}{c|}{Critical proportion} \\ \hline
 & \multicolumn{2}{c|}{Before variant} & \multicolumn{2}{c|}{50\% variant} & \multicolumn{2}{c|}{100\% variant} \\
 & Unvacc. & Vacc. & Unvacc. & Vacc. & Unvacc. & Vacc. \\ \hline
 donor area    & 0.51 & 0 & 0.68 & 0.10 & 0.76 & 0.33 \\ 
 nondonor area & 0.57 & 0 & 0.72 & 0.21 & 0.79 & 0.41 \\ \hline
\end{tabular}
\end{table}

\subsection{Optimization Results}
\label{sec:optimization}

For the scenarios in Table \ref{tab:scenarios}, the optimization found a policy that is better than any priority policy. The results are shown in Table \ref{tab:opt_results}. Area 1 is the donor; the nondonor areas 2, 3, and 4 are identical. The best policies found give the vaccine to the donor on the days listed, though on a few of these days it is split between areas. In Scenario 3.2, the best policy has 0.7\% fewer donor deaths than the best priority policy (where the donor has second priority) and 3.5\% fewer than the donor-first policy. The improvement is somewhat larger for Scenarios 3.1 and 4.1 and is negligible for Scenario 10.1. Although these improvements are modest, the best policy also reduces total deaths and delays the arrival of the variant. The best policy for Scenarios 3.1 and 4.1 has a switching form. 

\begin{figure}[!ht]
    \centering
    \includegraphics[width=4in]{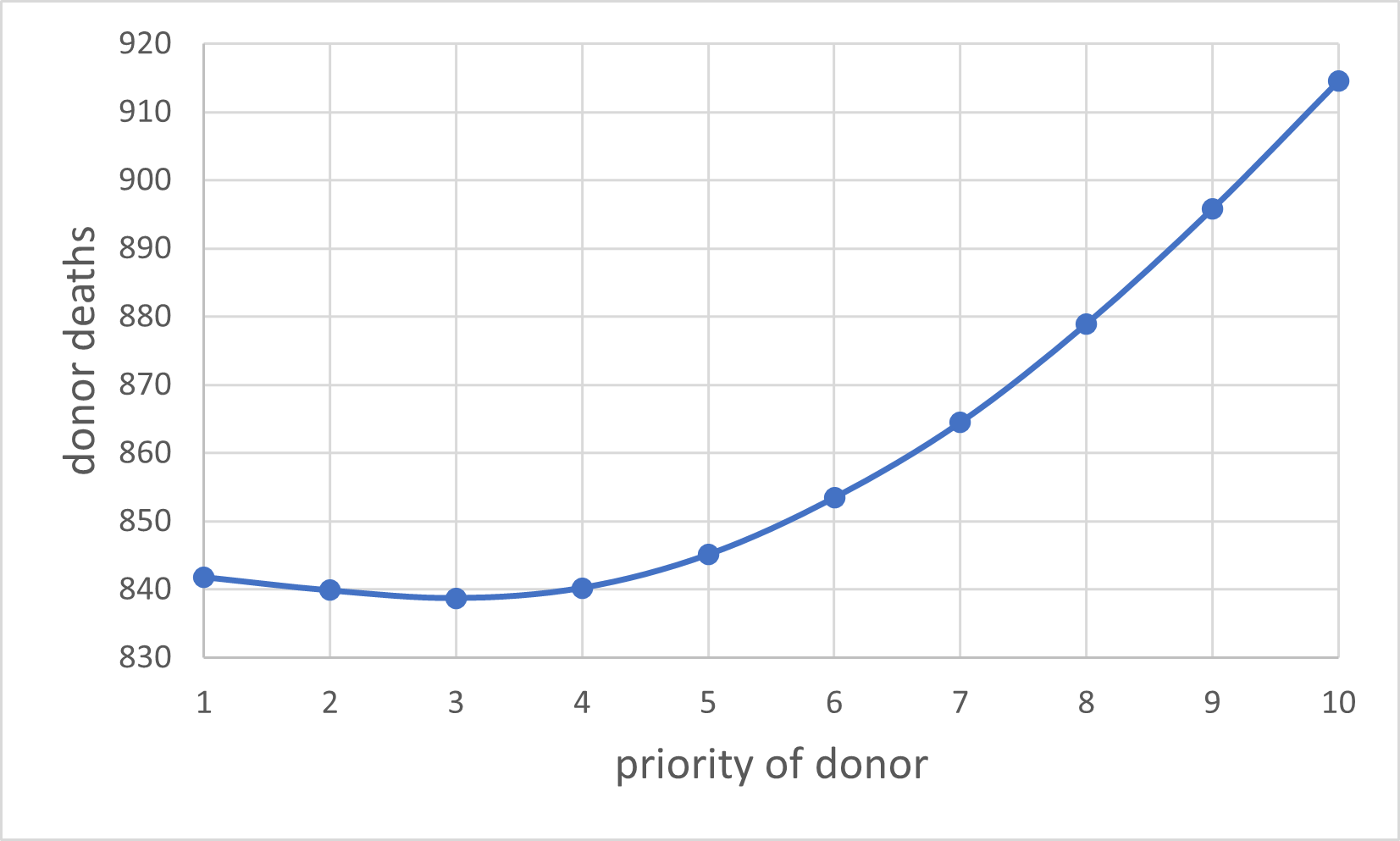}
    \caption{Effect of donor priority on deaths, Scenario 10.1}
    \label{fig:C10.1}
\end{figure}

In Scenario 10.1, area 1 is the donor and nondonor areas 2-10 have different initial infection rates, with area 2 having the greatest. The priority order of the nondonor areas is 2,...,10, i.e., the greatest infection rate is given highest priority.
The best policy found is nearly identical to the best priority policy, both in structure and in donor deaths. This is the priority policy shown in Table \ref{tab:opt_results}. It gives priority 3 to the donor. Donor deaths for other priority policies are graphed in Figure \ref{fig:C10.1}. Because of the large number of areas and possible forms of policy, we are less confident that the algorithm found a near-optimal policy. However, the number of possible priority policies is much larger and it makes sense that one of them is near-optimal.


These results demonstrate that optimal policies can use switching, not static priority. The more complex policies found can also be used to construct switching policies with fewer switch points, which are easier to implement, that also outperform priority policies.


\begin{table}[!h] 
\caption{Priority and best policies. The donor area receives vaccines on the days listed.}
\label{tab:opt_results}
\begin{tabular}{| l | l | c c | c | c |}
\hline 
 &  & \multicolumn{2}{c|}{Deaths } & Time of &  \\ 
Scenario & Policy & Donor & Total & variant (days) & Best $\lambda$ \\  \hline \hline
3.1 & priority 1,2,3 & 414.6 & 1028.2 & 49.0 & \\
    & priority 2,1,3 & 412.9 & 902.4 & 69.5 & \\
    & priority 2,3,1 & 417.7 & 739.1 & 165.1 & \\
    & days 36-78 & 402.3 & 813.8 & 93.8 &  $2.3 \times 10^{-5}$ \\ \hline
3.2 & priority 1,2,3 & 576.6 & 1301.4 & 45.5 & \\
    & priority 2,1,3 & 560.4 & 1200.0 & 61.8 & \\
    & priority 2,3,1 & 570.0 & 1112.1 & 104.2 & \\
    & days 33-73, 85 & 556.6 & 1180.02 & 71.4 &  $10^{-5}$ \\ \hline
4.1 & priority 1,2,3,4 & 560.3 & 1628.4 & 45.8 & \\
    & priority 2,1,3,4 & 538.2 & 1517.3 & 56.1 & \\
    & priority 2,3,1,4 & 518.8 & 1385.6 & 68.5 & \\
    & priority 2,3,4,1 & 521.0 & 1256.1 & 100.3 & \\
    & days 42-73, 76 & 510.0 & 1352.4 & 76.5 & $1.9 \times 10^{-5}$  \\ \hline
10.1 & priority 2,3,1,4,...,10 & 838.7 & 3810.2 & 61.4 & \\
    & days 28-48 & 838.3 & 3873.9 & 62.4 & $8.0 \times 10^{-6}$ \\ \hline
\end{tabular}
\end{table}

\subsection{Sensitivity Analysis}
\label{sec:sensitivity}

This section examines the sensitivity of the best policy and predicted deaths to several model parameters. Starting with Scenario 3.2 (because it showed the most benefit of policies that are not donor-first), parameters were varied one at a time. For each parameter value, all priority policies were simulated and the optimization algorithm was run. The best policy found is compared to a donor first policy in Table \ref{tab:sens}.

For the first sensitivity, we assumed that all areas contribute to emergence of the variant, not just the nondonor areas (\say{global variant}) and doubled $\mu$, since a larger population is contributing to the variant. The donor-first policy is much better than other priority policies, and all policies have very similar time until the variant may appear. Next, the time horizon was extended. For both 270 and 360 days, the best policy found vaccinates the donor last; before that, it has some switching between the nondonor areas. In contrast, for the baseline scenario, the best policy vaccinates the donor after partially vaccinating both nondonor areas. Generally, we found that a donor-last policy performs better when the time horizon extends well beyond the mean time when the variant appears, so that the variant has more impact.

When behavior dynamics are increased (by decreasing $I_{\rm{max}}$), the best policy vaccinates the donor slightly earlier than in the baseline scenario. Eliminating behavior dynamics ($I_{\rm{max}}$=0) leads to much higher deaths. The best policy, by far, is donor-first in this more severe scenario.

When the mean infectious days until the variant is increased, the best policy is a priority policy that gives second priority to the donor, beginning donor vaccinations nine days earlier than in the baseline scenario. When the mean infectious days until the variant is decreased to 40,000, this policy, which is fairly close to the baseline best policy, is again best. However, when the mean is decreased to 35,000, the donor-first policy is best, but only sightly better than the donor-second policy. These results show that the best policy is very sensitive to when the variant emerges and that a switching policy is only needed in a narrow range of parameter values.

\begin{table}[!h] 
\caption{Sensitivity analysis for policies and deaths}
\label{tab:sens}
\begin{tabular}{| l | l | l | cc | c | c |}
\hline 
& & Best Policy & \multicolumn{2}{c|}{Deaths} & Donor-first Policy & Variant \\
Parameter (baseline) & Value & Vacc. Donor & Donor & Total & Donor Deaths & (days) \\ 
    \hline \hline
Scenario 3.2 (used as baseline) & -- & days 33-73, 85 & 556.6 & 1180.02 & 576.6 & 71.4  \\ \hline
Global variant, infectious days & -- & priority 1,2,3 & 588.8 & 1267.6 & 588.8 & 52.7 \\
until variant $\mu = 100,000$ & & & & & & \\ \hline 
Time horizon $T$ (180 days) & 270 & days 47-86 & 666.9 & 1425.2 & 762.1 & 104.6 \\ 
                               & 360 & days 47-86 & 632.8 & 1333.4 & 728.8 & 104.6 \\ \hline
Behavior dynamics $I_{\rm{max}}$ (0.03) & 0.02 & days 25-69 & 425.8 & 911.0 & 437.9 & 82.7 \\
& 0 & priority 1,2,3 & 836.7 & 2049.7 & 836.7 & 30.3 \\ \hline
Infectious days until variant & 70,000 & priority 2,1,3 & 488.6 & 1060.4 & 510.1 & 87.8 \\ 
$\mu$ (50,000) & 40,000 & priority 2,1,3 & 604.4 & 1279.6 & 607.5 & 50.6 \\ 
               & 35,000 & priority 1,2,3 & 621.4 & 1377.7 & 621.4 & 35.9 \\  \hline
\end{tabular}
\end{table}

\section{Future Work and Conclusions}
\label{sec:conc}

We have developed an SEIR-embedded optimization framework that determines vaccine allocation policies during a pandemic between a donor nation and one or more recipient (nondonor) nations.  We present an iterative linear programming approximation approach to confirm those instances where a static priority policy is optimal and, when not optimal, to find switching policies that are superior.  Policies with fewer switching points, which are easier to implement, can be constructed from the best policy found. The optimization method was demonstrated for up to 10 geographic areas and should be tractable for significantly more. Thus, it can be used to study a global pandemic by aggregating non-donor countries into groups.

The optimization model identifies realistic scenarios in which the donor nation prefers to give away vaccines before vaccinating its own population in order to minimize local deaths. Additionally, we find that policies other than donor-first can significantly delay the emergence of a more-contagious variant compared to donor-first, allowing more time for the development of improved treatments against the virus. Moreover, in all scenarios studied, policies other than donor-first achieve dramatic reduction in total deaths with only a small increase (and occasionally even a decrease) in donor-country deaths. Thus, vaccine distribution is not a zero-sum game between donor and nondonor countries.

Our SEIR model and optimization framework permits a flexible weighting of multiple objectives to minimize COVID-19 deaths in the donor country and in total so that questions of equitable distribution of vaccines during a pandemic can be thoroughly examined.

Our model does not consider logistical delays in globally distributing and administering vaccinations; vaccine supply is assumed to be the key constraint. Further, we assume the timeline is short enough so that vaccine willingness does not change significantly and that immunity does not wane enough to need to administer additional vaccine doses. Thus, there is no incentive to stockpile vaccine for future needs. Although almost all parameters of the model were estimated using COVID-19 data, we generally did not use country-specific data. To use the model for specific donor countries and groups of interacting countries, country-specific population, infection rate, and transmission data could be used. We leave this to future work.  

\backmatter

%

%

\section*{Declarations}

The authors declare that they have no competing interests.

\bibliography{references}

\end{document}